\documentstyle[sprocl,psfig,12pt]{article}

\begin{document}
\sloppy

\title{
ONE-- AND TWO--NEUTRON CAPTURE REACTIONS OF LIGHT NUCLEI IN NUCLEAR
ASTROPHYSICS}                 

\author{        
H. Herndl, R. Hofinger and H. Oberhummer} 

\address{Institut f\"ur Kernphysik, TU Wien,\\
Wiedner Hauptstr. 8--10, A--1040 Vienna, Austria}

\maketitle\abstracts{We discuss models to calculate one-- and 
two--neutron capture reactions on light nuclei. These are applied
to calculate the reaction rates of
$^{15}$N(n,$\gamma$)$^{16}$N,
$^{16}$N(n,$\gamma$)$^{17}$N and
$^4$He(2n,$\gamma$)$^6$He. The possible astrophysical importance
is discussed.}

\section{Introduction}

Neutron capture reactions on light nuclei play a role in various
astrophysical scenarios. In the framework of Inhomogeneous
Big Bang Models a high neutron flux can bridge the mass 5 and
mass 8 gaps. Subsequent neutron capture reactions may trigger
a primordial r--process \cite{rau94}. Another site for neutron
capture reactions is the high--entropy bubble formed during a
type II supernova \cite{woo92,mey92,woo94}. 
Due to the photodisintegration at very high
temperatures an $\alpha$--rich environment is created. When the
temperature has dropped heavier elements can be built up mainly
by $\alpha$-- and neutron--capture reactions. Again a critical question is
how the mass 5 and mass 8 gaps can be bridged.

In Section 2 we describe our model to calculate one--neutron
capture reactions. This model is applied to the reactions
$^{15}$N(n,$\gamma$)$^{16}$N and
$^{16}$N(n,$\gamma$)$^{17}$N. In Section 3 we will describe the
theory of calculating a two--neutron capture reaction in a
three--body model. We calculate the reaction rate of
$^4$He(2n,$\gamma$)$^6$He and compare the result with other
works. In the final chapter we discuss and summarize our
results.

\section{One--Neutron Capture Reactions}

One--nucleon capture reactions on light nuclei are usually dominated
by direct capture (DC) to bound states and resonant capture to
single, isolated resonances above the threshold. The level density 
above the threshold is
normally low, i.e.~between 0--10 levels per MeV.
We calculate the cross section and reaction rate of the capture reaction
with a hybrid model.
The DC cross section is evaluated in a potential model.
For the resonances we use the Breit--Wigner formula. We will discuss
both contributions separately.

\subsection{Direct Capture}

The potential model is
described by different authors\,\cite{kim87,obe91,moh93}. We use real 
folding potentials as optical potentials\,\cite{obe91,kob84}.

The DC cross section of a transition to a bound state is determined
by the overlap of the scattering wave function, the bound--state wave
function and the electromagnetic transition--operator. In most cases
only E1--transitions need to be taken into account.

The total nonresonant cross
section $\sigma^{\rm DC}_{\rm tot}$ is determined 
by the direct capture transitions
$\sigma^{\rm DC}_i$ to all bound states multiplied with the single particle
spectroscopic factors $C^2 S_i$:
\begin{equation}
\label{NR}
\sigma^{\rm DC}_{\rm tot} = \sum_{i} \: (C^{2} S)_i \sigma^{\rm DC}_i \quad .
\end{equation}

The spectroscopic factors 
can be determined experimentally from other reactions,
e.g., the spectroscopic factor of a (n,$\gamma$)--reaction can be
obtained from the (d,p)--reaction. Alternatively the spectroscopic
factors can also be determined from shell--model calculations.

>From the direct capture cross section we can obtain the reaction
rate by integrating over a Maxwell--Boltzmann velocity distribution.
We parametrize the non--resonant contribution to the rate as
\begin{equation}
\label{eq-dcrate}
N_{\rm A} <\sigma v>_{\rm nr} = A + B T_9 - C T_9^D \quad 
{\rm cm}^3 {\rm mole}^{-1} {\rm s}^{-1} \quad , 
\end{equation}
where $T_9$ denotes the temperature in GK. The constant term
represents the s--wave capture contribution and the term
proportional to $T_9$ the p--wave capture. All other
contributions, i.e.~higher partial waves and deviations from the
conventional behaviours, are fitted in the third term.

\subsection{Single, Isolated Resonances}

The cross section of a single isolated resonance in neutron capture
M
processes s well described by
the Breit--Wigner formula~\cite{bre36,bla62}:
\begin{equation}
\label{BW}
\sigma_{\rm r}(E) = 
\frac{\pi \hbar^2}{2 \mu E}
\frac{\left(2J+1\right)}{2\left(2j_{\rm t}+1\right)} 
\frac{\Gamma_{\rm n} \Gamma_{\rm\gamma}}
{\left(E_{\rm r} - E\right)^2 + 
\left(\frac{\Gamma_{\rm tot}}{2}\right)^2}\quad ,
\end{equation}
where $J$ and $j_{\rm t}$ are the spins of the resonance level and the
target nucleus, respectively, $E_{\rm r}$ is the resonance energy.
The partial widths of the entrance and exit channels
are $\Gamma_{\rm n}$  and $\Gamma_{\rm\gamma}$, respectively.
The total width $\Gamma_{\rm tot}$
is the sum over the partial widths of all channels.
The neutron partial width $\Gamma_{\rm n}$ can be expressed in terms of the
single--particle spectroscopic factor $C^2 S$ and the single--particle width
$\Gamma_{\rm s.p.}$ of the resonance state~\cite{wie82,her95}
\begin{equation}
\label{SF}
\Gamma_{\rm n} = (C^2 S) \Gamma_{\rm s.p.} \quad .
\end{equation}
The single--particle width $\Gamma_{\rm s.p.}$ can be calculated
from the scattering phase shifts of a scattering potential with the
potential depth being determined by matching the resonance energy.

The gamma partial widths $\Gamma_{\rm\gamma}$ are calculated from 
the electromagnetic
reduced transition probabilities B($J_{i}\rightarrow J_{f}$;L) which
carry the nuclear structure information of the resonance states 
and the final bound states \cite{bru77}.
The reduced transition rates are computed within the framework of the
shell--model.

The resonant reaction rate for an isolated narrow resonance can be
expressed
as\,\cite{wie82,her95}
\begin{eqnarray}
\label{RRATE}
N_{\rm A} <\sigma v>_{\rm r} & = & 1.54 \times 10^5 \mu^{-3/2} T_9^{-3/2}
\nonumber\\
& \times &
\sum_i (\omega \gamma)_i \exp{(-11.605 E_i / T_9)} \hspace{3mm}
{\rm cm}^3 {\rm mole}^{-1} {\rm s}^{-1} ,
\end{eqnarray}
where the resonance strength $\omega \gamma$ is defined as
\begin{equation}
\omega \gamma = \frac {2J+1}{2(2j_t+1)} \frac{\Gamma_{\rm n}
\Gamma_{\rm\gamma}}{\Gamma_{\rm tot}} 
\end{equation}
and has to be inserted in eV in Eq.~\ref{RRATE}.
The resonance strength can be determined experimentally 
or derived from the calculated
partial widths.

\subsection{One--Neutron Capture on N--Isotopes}

We will now consider one--neutron capture reactions on neutron--rich
N--isotopes. We start with the reaction $^{15}$N(n,$\gamma$)$^{16}$N
which is known experimentally at stellar energies\,\cite{mei96}.
Therefore this reaction can be used as benchmark test for the validity
of our model.

The considered transitions for the direct capture of this reaction
are listed in
Table \ref{tab-dc}. The four lowest states of $^{16}$N can be
described by a coupling of the $1/2^-$ ground state of $^{15}$N with a
neutron from the 1d$_{5/2}$-- (resulting in the $2^-$ ground state and
a $3^-$ excited state) and the 2s$_{1/2}$--subshell (resulting in
the low--lying excited states $0^-$ and $1^-$). In a shell--model
description these states are good one--particle states. Therefore
the spectroscopic factors should be close to unity. In fact the 
shell--model calculations result in spectroscopic factors of 
about 0.9 for all
states \cite{mei96}. However, the spectroscopic factors from a
(d,p)--experiment are considerably lower by an average factor of
around 1.8 \cite{boh72}. The reason for this discrepancy is not known.
In our DC calculations we use the experimental spectroscopic factors.

All transitions listed in Table \ref{tab-dc} result from an
incoming p--wave.
The s--wave contribution which dominates at very low energies can
be obtained directly by extrapolating the thermal absorption cross
section with the $1/v$--law.

\begin{table}[t]
\caption{Considered transitions for the direct capture reactions on
N--isotopes.
Transitions with very small contributions are not included in the
table. The Q--values are in MeV.}
\vspace{0.2cm}
\begin{center}
\footnotesize
\begin{tabular}{|rrrrrr|}
\hline
reaction & Q--value & $J^{\pi}$ & $E_x$ (MeV) & transition & $C^2 S$ \\
\hline
$^{15}$N(n,$\gamma)^{16}$N & 2.491 & $2^-$ & 0.000 & p$\to$1d$_{5/2}$ &
0.550 \\
 & & $0^-$ & 0.120 & p$\to$2s$_{1/2}$ & 0.460 \\
 & & $3^-$ & 0.298 & p$\to$1d$_{5/2}$ & 0.540 \\
 & & $1^-$ & 0.397 & p$\to$2s$_{1/2}$ & 0.520 \\
$^{16}$N(n,$\gamma)^{17}$N & 5.883 & $1/2^-$ & 0.000 & p$\to$1d$_{5/2}$ &
0.589 \\
 & & $5/2^-$ & 1.907 & p$\to$1d$_{5/2}$ &
0.207 \\
 & & $7/2^-$ & 3.129 & p$\to$1d$_{5/2}$ & 1.457 \\
 & & $5/2^-$ & 4.415 & p$\to$2s$_{1/2}$ & 0.921 \\
\hline
\end{tabular}
\end{center}
\label{tab-dc}
\end{table}

\begin{table}[t]
\caption[resonance parameters 2]
{Adopted values for the resonance parameters for capture reactions 
on nitrogen isotopes.}
\begin{center}
\footnotesize
\begin{tabular}{|lcccrcc|}
\hline
reaction & $E_{\rm x}$ & $J^{\pi}$ & 
$E_{\rm res}$ & $\Gamma_{\rm n}\enspace\,$ & 
$\Gamma_{\gamma}$ & $\omega\gamma$\\
& $[$\footnotesize MeV$]$  & & $[$\footnotesize MeV$]$ & 
$[$\footnotesize eV$]\enspace\,$ & 
$[$\footnotesize eV$]$ & $[$\footnotesize eV$]$ \\ 
\hline
$^{15}\mbox{N(n,}\gamma\mbox{)}$$^{16}\mbox{N}\quad$ &
3.360 & $1^{+}$ & 0.862 & 15$\,$000 & 0.455 & 0.341  \\ 
$^{16}\mbox{N(n,}\gamma\mbox{)}$$^{17}\mbox{N}\quad$ &
5.904 & $7/2^{-}$ & 0.021 & 0.032 & 4.80$\,\cdot\,10^{-2}$ & 
0.015 \\ 
& 6.121 & $5/2^{+}$ & 0.238 & 1.2 & 4.80$\,\cdot\,10^{-2}$& 0.027\\
& 6.325 & $3/2^{+}$ & 0.442 & 20 & 5.46$\,\cdot\,10^{-2}$& 0.022\\
& 6.372 & $7/2^{+}$ & 0.489 & 20 & 1.52$\,\cdot\,10^{-2}$& 0.012\\
& 6.373 & $5/2^{+}$ & 0.490 & 600 & 0.110 & 0.066\\
& 6.470 & $1/2^{+}$ & 0.587 & 1$\,$750 & 2.510 & 0.501\\
& 6.685 & $3/2^{-}$ & 0.802 & 12$\,$500 & 5.660 & 2.263\\
& 6.737 & $7/2^{+}$ & 0.854 & 70 & 4.17$\,\cdot\,10^{-2}$&  0.033\\
& 6.835 & $3/2^{+}$ & 0.952 & 360 & 0.478& 0.191\\
\hline
\end{tabular}
\end{center}
\label{tab-res}
\end{table}

The resonance parameters are listed in Table \ref{tab-res}.
The total width of the 862\,keV resonance
of $\Gamma = 15$\,keV\,\cite{ajz86} corresponds to the neutron width 
of the state. The $\gamma$ width was estimated with the recommended
upper limit as $\Gamma_{\gamma} = 4.2$\,eV \cite{mei96}.
>From the shell model calculation we obtain a width
$\Gamma_{\gamma} = 0.455$\,eV which reduces the resonance strength
by about one order of magnitude. However, this resonance only has a
small influence on the reaction rate at very high temperatures.
Therefore this change of the resonance strength barely changes the
reaction rate.

\begin{table}[htb]
\caption[direct capture parameters]
{Parameters for the direct--capture contribution to the reaction rate.}
\begin{center}
\begin{tabular}{|lcrrr|}
\hline
 & \multicolumn{1}{c}{$A$} & \multicolumn{1}{c}{$B$} &
\multicolumn{1}{c}{$C$} & \multicolumn{1}{c|}{$D$} \\ \hline
$^{15}$N(n,$\gamma$)$^{16}$N & 3.18 & $3783.4$ & $335.2$ & $1.716$ \\
$^{16}$N(n,$\gamma$)$^{17}$N & & $3649.9$ & $437.5$ & $1.633$ \\
\hline
\end{tabular}
\end{center}
\label{tab-rates}
\end{table}

The parameters of the direct capture reaction rate (see Eq.~\ref{eq-dcrate})
are listed in Table \ref{tab-rates}.
The resulting reaction rate agrees very well with the rate of
Meissner {\it et al}.~\cite{mei96} The rate is clearly dominated by the direct
capture contribution. Both rates show good agreement with the
experimental data. For a discussion of the experimental data we
refer the reader to the paper of Meissner\,\cite{mei96}.

The bound levels of $^{17}$N are known from experiment. We
used spectroscopic factors calculated in the shell--model
for the DC calculation. The important DC transitions are
listed in Table \ref{tab-dc}. Several levels are known above
the threshold but without spin/parity assignments. It is
impossible to assign the levels with the help of our shell
model calculation. Therefore we use the shell--model energies for the
resonances. The adopted resonance parameters are shown in Table \ref{tab-res}.
The two resonances at 21\,keV and 238\,keV are the main
contributions to the reaction rate. 

Since there is no s--wave transition the parameter $A$ of the
reaction rate vanishes (see Table \ref{tab-rates}).
We find a considerable enhancement of our rate compared to
the calculation of Rauscher {\it et al}.~\cite{rau94} In that work the
resonance at 197\,keV was given a hypothetical $5/2^+$
assignment. Since no resonance at lower energies was
included, our 21\,keV resonance causes a strong increase
of the rate at low temperatures.

\section{Two--Neutron Capture Reactions}\label{results for
two-neutron capture reactions}

In this section we will discuss the possibilities to calculate
a three--body reaction rate. We will apply the theory to the reaction
$^4{\rm He}(2{\rm n},\gamma)^6{\rm He}$.

The reaction rate of
$^4{\rm He}(2{\rm n},\gamma)^6{\rm He}$ can be calculated
as a sequential two--step
process.
We call this method the
$^6{\rm He}\equiv\alpha$+n+n approach.
Here it is assumed that in a first step the unstable
nucleus $^5{\rm He}$ is formed via the reaction
$^4{\rm He}+n\rightarrow\,^5{\rm He}$ (negative Q--value $Q_1$).
The second step $^5{\rm He}+n\rightarrow\,^6{\rm He}+\gamma$
(positive Q--value $Q_2$)
is treated as a two--body problem in the initial and as a three--body problem
in the final state. 
The values for spin--parities, $Q$--values and widths used 
in our calculations are given in Table \ref{tab3body-he4-1}.
The total Q--value $Q_{12}$ is the sum of the Q--values $Q_1$ and
$Q_2$.

\begin{table}[htb]
\caption[Spin-parities, $Q$-values and widths for 
$^4{\rm He}(2n,\gamma)^6{\rm He}$]
{\label{tab3body-he4-1}Spin/parities, $Q$--values and widths (in MeV) used in
our $\alpha$+n+n calculations of the reaction rate for
the reaction $^4{\rm He}(2n,\gamma)^6{\rm He}$.} 
\begin{center}
\footnotesize
\begin{tabular}{|ccccccc|}
\hline 
$J^\pi(^4{\rm He})$ & $J^\pi(^5{\rm He})$ &
$J^\pi(^6{\rm He})$ & 
$Q_{1}$ & $Q_{2}$ &
$Q_{12}$ &$\Gamma_c$ \\ \hline
$0^+$ & $3/2^-$ & $0^+$ & $-0.89$ & 
$1.87$ & $0.98$ & $0.76$ \\ \hline
\end{tabular}
\end{center}
\end{table}

In this two--step model the reaction rate of $^4$He(2n,$\gamma$)$^6$He 
is given by
\begin{equation}
N_{\rm He}^2\left\langle 2{\rm n}  ^4{\rm He}\right\rangle=
2 \int \int dE_1 dE_2 \frac{\hbar}{\Gamma (E_1)}
\frac{d<{\rm n} ^4{\rm He}>}{dE_1} \frac{d<{\rm n} ^5{\rm He}>}{dE_2} \quad .
\end{equation}
The energies $E_1$ and $E_2$ denote the center--of--mass collision
energies of the first (second) neutron with the $^4$He ($^5$He)
nucleus. The width of the intermediate state, $\Gamma (E_1)$,
is energy--dependent.
The differential rates of the two steps represent
the product of the cross section with the Maxwell--Boltzmann
distribution:
\begin{equation}
\frac{d<\sigma v>}{dE} = \left( \frac{8}{\pi \mu} \right)
\left( \frac{1}{kT} \right)^{3/2} \sigma (E) E 
\exp{\left( - \frac{E}{kT} \right)} \quad .
\end{equation}
The whole procedure of calculating the cross section
is described in detail by Efros {\it et al}~\cite{efros96} and by
Balogh\,\cite{balogh97}.

The numerical results for the reaction rate $N_{\rm A}^2\left\langle
2{\rm n} ^4{\rm He}\right\rangle$ calculated in this model
are fitted to the expression
\begin{equation}\label{eq3body-5}
N_{\rm A}^2\left\langle 2{\rm n} ^4{\rm He}\right\rangle=
\sum_{i} a_i T_9^i\,{\rm cm}^6\,{\rm s}^{-1}\,{\rm mole}^{-2}
\end{equation}
in the temperature regions $0.1\le T_9 < 1$ and $1\le T_9 \le 3$.
>From a comparison with the situation of small resonances
it should be possible to perform a fit to a function
$\sim a_1 T_9^{a_2} \exp(-a_3/T_9)$, but it turned out that such a
fit leads only for $T_9\ge 1$ to useful results.
Therefore, we decided to use the parametrization of Eq.~(\ref{eq3body-5}),
which reproduces the calculated data with an error less than $0.25\%$ 
in the whole temperature range.
The parameters are given in Table \ref{tab3body-1}.

\begin{table}[htb]
\caption 
{Fit--parameters $a_i$ for the reaction rate
$N_{\rm A}^2\left\langle 2n ^4{\rm He}\right\rangle$
calculated in our $^6{\rm He}\equiv \alpha+$n+n approach
and from the E1--strength obtained in the work of Danilin.
\label{tab3body-1}}
\begin{center}
\footnotesize
\begin{tabular}{|crrrr|}
\hline
& \multicolumn{2}{c}{$\alpha+n+n$} 
& \multicolumn{2}{c|}{Danilin {\it et al}} \\     
     & \multicolumn{1}{c}{$0.1\le T_9 < 1$} & 
       \multicolumn{1}{c}{$1  \le T_9 \le 3$}
     & \multicolumn{1}{c}{$0.1\le T_9 < 1$} & 
       \multicolumn{1}{c|}{$1  \le T_9 \le 3$} \\ \hline
 $a_0$ &  $ 2.75605{\rm E\,-16}$ & $-4.31090{\rm E\,-11}$ & $-4.94120{\rm E\,-11}$ & $ 2.27413{\rm E\,-10}$ \\
 $a_1$ &  $ 2.88001{\rm E\,-12}$ & $ 1.95298{\rm E\,-10}$ & $ 2.86628{\rm E\,-09}$ & $-2.40420{\rm E\,-10}$ \\
 $a_2$ &  $ 4.41655{\rm E\,-12}$ & $-3.29920{\rm E\,-10}$ & $-4.90850{\rm E\,-08}$ & $ 5.64850{\rm E\,-09}$ \\
 $a_3$ &  $-1.35270{\rm E\,-12}$ & $ 2.77932{\rm E\,-10}$ & $ 3.46189{\rm E\,-07}$ & $-2.23220{\rm E\,-09}$ \\
 $a_4$ &  $ 1.68846{\rm E\,-11}$ & $-1.02300{\rm E\,-10}$ & $-1.08100{\rm E\,-06}$ & $ 3.69620{\rm E\,-10}$ \\
 $a_5$ &  $-2.45810{\rm E\,-11}$ & $ 1.80468{\rm E\,-11}$ & $ 1.87337{\rm E\,-06}$ & $-2.62480{\rm E\,-11}$ \\
 $a_6$ &  $ 1.41336{\rm E\,-11}$ & $-1.26680{\rm E\,-12}$ & $-1.85880{\rm E\,-06}$ & $4.500240{\rm E\,-13}$  \\
 $a_7$ &  $ 2.08214{\rm E\,-11}$ & \multicolumn{1}{c}{---}    & $ 9.89903{\rm E\,-07}$ &  \multicolumn{1}{c|}{---}     \\
 $a_8$ &  $-2.67550{\rm E\,-11}$ & \multicolumn{1}{c}{---}    & $-2.19670{\rm E\,-07}$ &  \multicolumn{1}{c|}{---}     \\
 $a_9$ &  $ 8.22640{\rm E\,-12}$ & \multicolumn{1}{c}{---}    & \multicolumn{1}{c}{---}&  \multicolumn{1}{c|}{---}     \\  
\hline
\end{tabular}
\end{center}
\end{table}

In the work of Danilin {\it et al}~\cite{danilin93}
the wave functions for ground and scattering states of the
halo nucleus $^6{\rm He}$ are calculated in an $\alpha$+n+n
three-body model and are used to predict the strengths of nuclear
and electric dipole excitations. From the astrophysical point
of view we are especially interested in the electric dipole
strength function $dB^{\rm E1}/dE_{\gamma}$,
which is directly connected to the cross section
for photodisintegration $\sigma_{\gamma}^{\rm E1}$ of the nucleus
$^6{\rm He}$ into $^4{\rm He}$ and two neutrons by \cite{sackett93}
\begin{equation}\label{eq3body-da1}
\sigma_{\gamma}^{\rm E1}(E_{\gamma})=\frac{16\pi^3}{9}
\frac{E_{\gamma}}{\hbar c}\frac{dB^{\rm E1}}{dE_{\gamma}},
\end{equation}
where $E_\gamma=E+S_{\rm 2n}$ is the photon energy, which can be written
as the sum of the two--neutron energy $E=E_1+E_2$ and the two--neutron
separation energy, $S_{\rm 2n}\equiv Q_{12}$. We refer the reader to
the work of Danilin {\it et al}~\cite{danilin93} about details of the
calculation.

\begin{figure}[t]
\centerline{\psfig{file=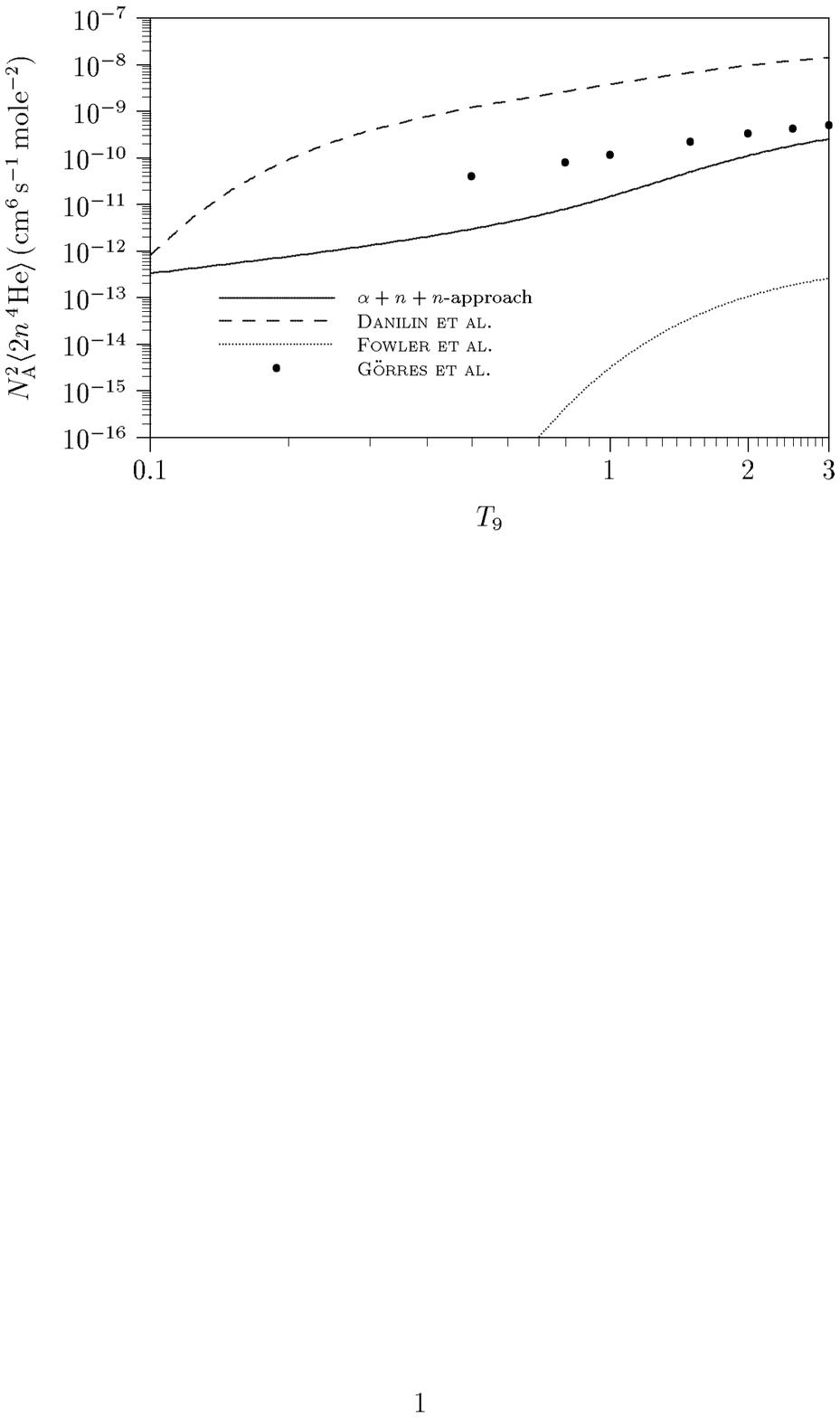,clip=}}
\caption[Comparison of different reaction rates 
$N_{\rm A}^2\left\langle 2n^4{\rm He}
\right\rangle$]
{\label{fig3body-he4-1}
Comparision of different rates 
$N_{\rm A}^2\left\langle 2{\rm n}^4{\rm He} 
\right\rangle$ for the formation of $^6$.}
\end{figure}

In Fig.~\ref{fig3body-he4-1} the reaction rates
$N_{\rm A}^2\left\langle 2{\rm n} ^4{\rm He}\right\rangle$
for the formation of $^6{\rm He}$ calculated 
in the $\alpha$+n+n approach and deduced from the E1--strength
function of Danilin {\it et al}~\cite{danilin93,thompson97}
are depicted. Furthermore, the results obtained in previous 
studies of Fowler {\it et al}~\cite{fowler75}
and G\"orres {\it et al}~\cite{goerres95} are shown (we have
used the constructive rate of
that publication).
Our results obtained in the $\alpha$+n+n approach
show a relatively good agreement
with those obtained by G\"orres \cite{goerres95}, who also used
a two--step model with a simple direct capture calculation
for the second step.
Below $T_9=0.8$ no data for the rate $N_{\rm A}^2\left\langle
2{\rm n}^4{\rm He}\right\rangle$ are available in the work 
of G\"orres {\it et al}~\cite{goerres95}.
Our calculations are larger by more than three orders of magnitude
compared to the calculations of Fowler {\it et al}~\cite{fowler75}.
This is due to the fact that in the work of Fowler
it was assumed that
the second step ($^5{\rm He}+n\rightarrow\,^6{\rm He}+\gamma$)
proceeds via a resonant state in $^6{\rm He}$ near the threshold
followed by the emission of E2--radiation,
which is much more less likely to take place than the nonresonant
E1 direct capture of a neutron into the ground state of $^6{\rm He}$.

The most remarkable fact is that the rate deduced from the
E1--strength function of Danilin {\it et al}~\cite{danilin93,thompson97}
is for $T_9 > 0.2$ approximately two orders of magnitude larger than our
results obtained in the $\alpha$+n+n approach.
A possible explanation of this enhancement is that in our 
$\alpha$+n+n calculations
only the sequential process is taken into account
whereas in the calculations of Danilin {\it et al}~\cite{danilin93}
the simultaneous decay ($^6{\rm He}+\gamma\rightarrow
\,^4$+n+n) is also included.
Furthermore, it is not possible to describe the threshold behaviour
of the cross section $\sigma_\gamma^{\rm E1}$ (with the help
of the reciprocity theorem) very well for broad resonances
in a two--step model.

\section{Summary and Discussion}

We calculated reaction rates for one--neutron capture on $^{15}$N
and $^{16}$N and two--neutron capture on $^4$He. The rate for
$^{15}$N(n,$\gamma$)$^{16}$N is in good agreement with both
previous calculations and experimental data. In general the
neutron capture rates on stable targets are quite well known.
The reaction $^{16}$N(n,$\gamma$)$^{17}$N is an example of a
capture reaction on an unstable target. We find a considerable
enhancement to a previous calculation. This enhancement can be
also be observed at various other capture reactions on unstable
targets in this mass range \cite{her98}. This fact could
influence the reaction path both in the nucleosynthesis of
Inhomogeneous Big Bang Models and in the alpha--rich freeze--out
of type II supernovae.

For the reaction $^4$He(2n,$\gamma$)$^6$He we also find a strong
enhancement of the rate compared to the calculation of
Fowler \cite{fowler75}. But from the calculation of the inverse
photodisintegration rate by Danilin {\it et al}~\cite{danilin93} we
deduce an even much higher rate. This is probably due to the fact that
this rate also includes the simultaneous capture of two neutrons.
But even with this enhanced rate the path via
$^4$He(2n,$\gamma$)$^6$He
is dominated by the reaction
$^4$He($\alpha$n,$\gamma$)$^9$Be at conditions typical for the
alpha--rich freeze--out. This is due to the effective destruction
of $^6$He through photodisintegration.
\section*{References}
\end{document}